\documentclass[aps,prl,twocolumn,superscriptaddress,showpacs]{revtex4}

\usepackage{color}
\usepackage{graphicx}

\newcommand{\gp}{$\vec \gamma p \rightarrow p\pi^+\pi^-${ }}

\begin{document}

\title{Beam-Helicity Asymmetries in Double-Charged-Pion 
Photoproduction on the Proton}

\newcommand*{\ASU}{Arizona State University, Tempe, Arizona 85287-1504}
\affiliation{\ASU}
\newcommand*{\UCLA}{University of California at Los Angeles, Los Angeles,
 California  90095-1547}
\affiliation{\UCLA}
\newcommand*{\CMU}{Carnegie Mellon University, Pittsburgh, 
Pennsylvania 15213}
\affiliation{\CMU}
\newcommand*{\CUA}{Catholic University of America, Washington, 
D.C. 20064}
\affiliation{\CUA}
\newcommand*{\SACLAY}{CEA-Saclay, Service de Physique Nucl\'eaire, 
F91191 Gif-sur-Yvette,Cedex, France}
\affiliation{\SACLAY}
\newcommand*{\CNU}{Christopher Newport University, Newport News, 
Virginia 23606}
\affiliation{\CNU}
\newcommand*{\UCONN}{University of Connecticut, Storrs, 
Connecticut 06269}
\affiliation{\UCONN}
\newcommand*{\ECOSSEE}{Edinburgh University, Edinburgh EH9 3JZ, 
United Kingdom}
\affiliation{\ECOSSEE}
\newcommand*{\FIU}{Florida International University, Miami, 
Florida 33199}
\affiliation{\FIU}
\newcommand*{\FSU}{Florida State University, Tallahassee, Florida 32306}
\affiliation{\FSU}
\newcommand*{\GWU}{The George Washington University, 
Washington, DC 20052}
\affiliation{\GWU}
\newcommand*{\ECOSSEG}{University of Glasgow, Glasgow G12 8QQ, 
United Kingdom}
\affiliation{\ECOSSEG}
\newcommand*{\ISU}{Idaho State University, Pocatello, Idaho 83209}
\affiliation{\ISU}
\newcommand*{\INFNFR}{INFN, Laboratori Nazionali di Frascati, 
Frascati, Italy}
\affiliation{\INFNFR}
\newcommand*{\INFNGE}{INFN, Sezione di Genova, 16146 Genova, Italy}
\affiliation{\INFNGE}
\newcommand*{\ORSAY}{Institut de Physique Nucleaire ORSAY, Orsay, France}
\affiliation{\ORSAY}
\newcommand*{\ITEP}{Institute of Theoretical and Experimental Physics, 
Moscow, 117259, Russia}
\affiliation{\ITEP}
\newcommand*{\JMU}{James Madison University, Harrisonburg, 
Virginia 22807}
\affiliation{\JMU}
\newcommand*{\KYUNGPOOK}{Kyungpook National University, 
Daegu 702-701, South Korea}
\affiliation{\KYUNGPOOK}
\newcommand*{\MIT}{Massachusetts Institute of Technology, Cambridge, 
Massachusetts  02139-4307}
\affiliation{\MIT}
\newcommand*{\UMASS}{University of Massachusetts, Amherst, 
Massachusetts  01003}
\affiliation{\UMASS}
\newcommand*{\MAINZ }{Institut f\"ur Kernphysik, Johannes 
Gutenberg-Universit\"at Mainz, 55099 Mainz, Germany}
\affiliation{\MAINZ } 
\newcommand*{\MOSCOW}{Moscow State University, Skobeltsyn Nuclear Physics
Institute, 119899 Moscow, Russia}
\affiliation{\MOSCOW}
\newcommand*{\UNH}{University of New Hampshire, Durham, 
New Hampshire 03824-3568}
\affiliation{\UNH}
\newcommand*{\NSU}{Norfolk State University, Norfolk, Virginia 23504}
\affiliation{\NSU}
\newcommand*{\OHIOU}{Ohio University, Athens, Ohio  45701}
\affiliation{\OHIOU}
\newcommand*{\ODU}{Old Dominion University, Norfolk, Virginia 23529}
\affiliation{\ODU}
\newcommand*{\PITT}{University of Pittsburgh, Pittsburgh, 
Pennsylvania 15260}
\affiliation{\PITT}
\newcommand*{\RPI}{Rensselaer Polytechnic Institute, Troy, 
New York 12180-3590}
\affiliation{\RPI}
\newcommand*{\RICE}{Rice University, Houston, Texas 77005-1892}
\affiliation{\RICE}
\newcommand*{\URICH}{University of Richmond, Richmond, Virginia 23173}
\affiliation{\URICH}
\newcommand*{\SCAROLINA}{University of South Carolina, Columbia, 
South Carolina 29208}
\affiliation{\SCAROLINA}
\newcommand*{\JLAB}{Thomas Jefferson National Accelerator Facility, 
Newport News, Virginia 23606}
\affiliation{\JLAB}
\newcommand*{\UNIONC}{Union College, Schenectady, NY 12308}
\affiliation{\UNIONC}
\newcommand*{\VT}{Virginia Polytechnic Institute and State University, 
Blacksburg, Virginia   24061-0435}
\affiliation{\VT}
\newcommand*{\VIRGINIA}{University of Virginia, Charlottesville, 
Virginia 22901}
\affiliation{\VIRGINIA}
\newcommand*{\WM}{College of William and Mary, Williamsburg, 
Virginia 23187-8795}
\affiliation{\WM}
\newcommand*{\YEREVAN}{Yerevan Physics Institute, 375036 Yerevan, 
Armenia}
\affiliation{\YEREVAN}
\newcommand*{\deceased}{Deceased}
\newcommand*{\NOWOHIOU}{Ohio University, Athens, Ohio  45701}
\newcommand*{\NOWINDSTRA}{Systems Planning and Analysis, Alexandria, 
Virginia 22311}
\newcommand*{\NOWUNH}{University of New Hampshire, Durham, 
New Hampshire 03824-3568}
\newcommand*{\NOWCUA}{Catholic University of America, 
Washington, D.C. 20064}
\newcommand*{\NOWSCAROLINA}{University of South Carolina, 
Columbia, South Carolina 29208}
\newcommand*{\NOWUMASS}{University of Massachusetts, Amherst, 
Massachusetts  01003}
\newcommand*{\NOWMIT}{Massachusetts Institute of Technology, 
Cambridge, Massachusetts  02139-4307}
\newcommand*{\NOWODU}{Old Dominion University, Norfolk, Virginia 23529}
\newcommand*{\NOWGEISSEN}{Physikalisches Institut der Universit\"at 
Gie{\ss}en, 35392 Gie{\ss}en, Germany}
\newcommand*{\NOWNONE}{unknown, }

\author {S.~Strauch} 
\affiliation{\GWU}
\affiliation{\SCAROLINA}
\author {B.L.~Berman} 
\affiliation{\GWU}

\author {G.~Adams} 
\affiliation{\RPI}
\author {P.~Ambrozewicz} 
\affiliation{\FIU}
\author {M.~Anghinolfi} 
\affiliation{\INFNGE}
\author {B.~Asavapibhop} 
\affiliation{\UMASS}
\author {G.~Asryan} 
\affiliation{\YEREVAN}
\author {G.~Audit} 
\affiliation{\SACLAY}
\author {H.~Avakian} 
\affiliation{\INFNFR}
\affiliation{\JLAB}

\author {H.~Bagdasaryan} 
\affiliation{\ODU}
% ...
\collaboration{The CLAS Collaboration}
     \noaffiliation

\date{July 29, 2005}

\begin{abstract}
Beam-helicity asymmetries for the two-pion-photoproduction reaction
\gp have been studied for the first time in the resonance region for
center-of-mass energies between 1.35~GeV and 2.30~GeV.  The experiment
was performed at Jefferson Lab with the CEBAF Large Acceptance
Spectrometer using circularly polarized tagged photons incident on an
unpolarized hydrogen target.  Beam-helicity-dependent angular
distributions of the final-state particles were measured. The large
cross-section asymmetries exhibit strong sensitivity to the kinematics
and dynamics of the reaction. The data are compared with the results
of various phenomenological model calculations, and show that these
models currently do not provide an adequate description for the
behavior of this new observable.
\end{abstract}

\maketitle

%
%
The study of the baryon spectrum provides an avenue to a deeper
understanding of the strong interaction, since the properties of the
excited states of baryons reflect the dynamics and relevant degrees of
freedom within them. Many nucleon resonances in the mass region above
1.6~GeV decay predominantly through either $\pi\Delta$ or $\rho N$
intermediate states into $\pi\pi N$ final states (see the Particle-Data
Group review \cite{pdg04}). Resonances predicted by symmetric
quark models, but not observed in the $\pi N$ channel (the so-called
``missing'' resonances), are predicted to lie in the region of
$W>1.8$~GeV \cite{capstick94}. This makes electromagnetic
double-pion production an important tool in the investigation of the
structure of the nucleon.

To date, a rather large amount of unpolarized cross-section
measurements of double-pion photo- and electroproduction on the proton
have been reported by several collaborations \cite{ABBHHM68, ABBHHM69,
Braghieri95, Haerter97, Wolf00, Langgaertner01, Assafiri03, Ripani03,
Philips03, Bellis04}. However, the database collected for
polarization observables remains quite sparse.  Polarization degrees
of freedom in charged double-pion production have been studied at SLAC
\cite{Ballam72} and in the context of the Gerasimov-Drell-Hearn sum
rule at MAMI \cite{Ahrens03}.

On the theoretical side, some experience has been gained during the
last decade \cite{gomez-tejedor96,nacher01,nacher02,Roca04,
mokeev01,burkert04,mokeev04,Roberts97,Fix04}. It should be noted that
the various models which are presently used are constructed according
to the same scheme --- effective Lagrangian densities, where the
parameters for resonant and background mechanisms (contact and $u$,
$t$-channel pole terms) are either taken 
from other experiments or are treated as free parameters in the
analysis. Aside from the wide variations in the corresponding coupling
constants allowed by the Particle-Data Group listing, the primary
source of differences between the models is the treatment of the
background, which appears to be very complicated in the effective
Lagrangian approach for double-pion photoproduction.  A better
understanding of the double-pion photoproduction dynamics is vital for
the reliable extraction of $N^*$ photocouplings. Polarization data,
being particularly sensitive to interference effects, are expected to
provide valuable constraints.

In this Letter, we report the first comprehensive measurement of the
beam-helicity asymmetry \cite{Roberts04}
\begin{equation}
   I^\odot = \frac{1}{P_\gamma} \cdot 
   \frac{\sigma^+ - \sigma^-}{\sigma^+ + \sigma^-}
\end{equation}
in the \gp reaction, for energies $W$ between 1.35~GeV and 2.30~GeV in
the center of mass, where the photon beam is circularly polarized and
neither target nor recoil polarization is specified. $P_\gamma$ is the
degree of circular polarization of the photon and $\sigma^\pm$ are the
cross sections for the two photon-helicity states $\lambda_\gamma=\pm
1$.  Here, we give a brief overview of our data and demonstrate, by
means of a phenomenological model, the sensitivity of this observable
to the dynamics of the reaction.

%
%
The experiment was performed in Hall B at the Thomas Jefferson
National Accelerator Facility (Jefferson Lab).  Longitudinally
polarized electrons with an energy $E_0=2.445$~GeV 
were incident on a thin radiator. The beam polarization was routinely
monitored during data taking by a M{\o}ller polarimeter and was, on
average, 0.67. A photon tagger system \cite{Sober00} was used to tag
photons in the energy range between 0.5~GeV and 2.3~GeV, with an
energy resolution of 0.1\% $E_0$.  The degree of circular polarization
of the photon beam is proportional to the electron-beam polarization
and is a monotonic function of the ratio of the photon and incident
electron energies \cite{Maximon59}. The degree of photon-beam
polarization varied from $\approx 0.16$ at the lowest photon energy up
to $\approx 0.66$ at the highest energy. The photon-helicity state
changes with the electron-beam helicity, which was flipped
pseudo-randomly at a rate of 30~Hz. The collimated photon beam
irradiated an 18-cm thick liquid-hydrogen target.  The final-state
hadrons were detected in the CEBAF Large Acceptance Spectrometer
(CLAS) \cite{Mecking03}.  The CLAS provides a large coverage for
charged particles that includes particle momenta down to 0.25~GeV/c
and polar angles in the range $8^\circ < \theta_{\rm lab} <
145^\circ$.  The event trigger required a coincidence between a
scattered-electron signal from the photon tagger and at least one
charged track in the CLAS. The four-momentum vectors of the particles
were reconstructed from their tracks in the toroidal magnetic field of
the spectrometer by a set of three drift-chamber packages and by
particle identification using time-of-flight information from plastic
scintillators located about 5~m from the target.

%
%
The \gp reaction channel was identified in this kinematically complete
experiment by the missing-mass technique, requiring either the
detection of all three final-state particles
or the detection of two out of the three particles.

\begin{figure}[!ht]
\includegraphics[width=\columnwidth]{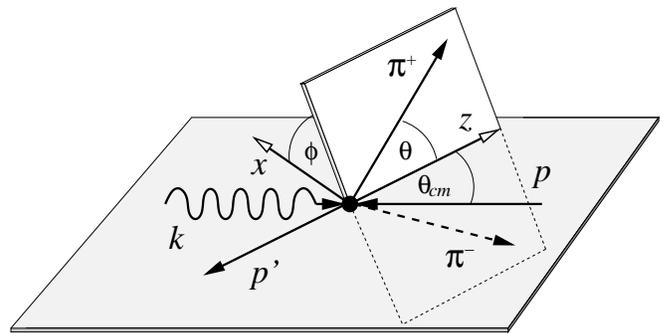}
\caption{\label{fig:def} Angle definitions for the circularly
polarized real-photon reaction \gp; $\theta_{cm}$ is defined in the
overall center-of-mass frame, and $\theta$ and $\phi$ are defined as
the $\pi^+$ polar and azimuthal angles in the rest frame of the
$\pi^+\pi^-$ system with the $z$ direction along the total
momentum of the $\pi^+\pi^-$ system (helicity frame).}
\end{figure}

A schematic view of the reaction, together with angle definitions, is
shown in Fig.~\ref{fig:def}.
A total of $3 \times 10^7$ $p\pi^+\pi^-$ events were accumulated for
both helicity states $N^\pm$.
Experimental values of the helicity asymmetry were then obtained as
\begin{equation}
   I^\odot_{\rm exp} = \frac{1}{\rule[2mm]{0mm}{1.5mm}\bar P_\gamma} \cdot
     \frac{N^+/\alpha^+ - N^-/\alpha^-}{N^+/\alpha^+ +
     N^-/\alpha^-}\;,
\end{equation}
where $\alpha^\pm = \frac{1}{2}(1 \pm a_c)$ accounts for
helicity-dependent differences in the luminosity due to a small
electron-beam-charge asymmetry $a_c \approx -0.0044$.  The value of
$a_c$ was determined from helicity asymmetries in single-pion
photoproduction for data that were obtained simultaneously with the
double-pion photoproduction data. Any observed asymmetry in this
reaction is instrumental \cite{Barker:1975bp}.
The experimental asymmetries have not been corrected for the CLAS
acceptance. In order to allow for an analysis as model-independent as
possible, the data are compared with event-weighted mean values of
asymmetries from model calculations \cite{method}. The determination
of these mean values does not require a knowledge of the CLAS
acceptance. The only major source of systematic uncertainty is the
value for the beam polarization, which is known to about 3\%. The
uncertainty from the beam-charge asymmetry is negligible (less than
$10^{-3}$).

\begin{figure}[!ht]
\includegraphics[width=\columnwidth]{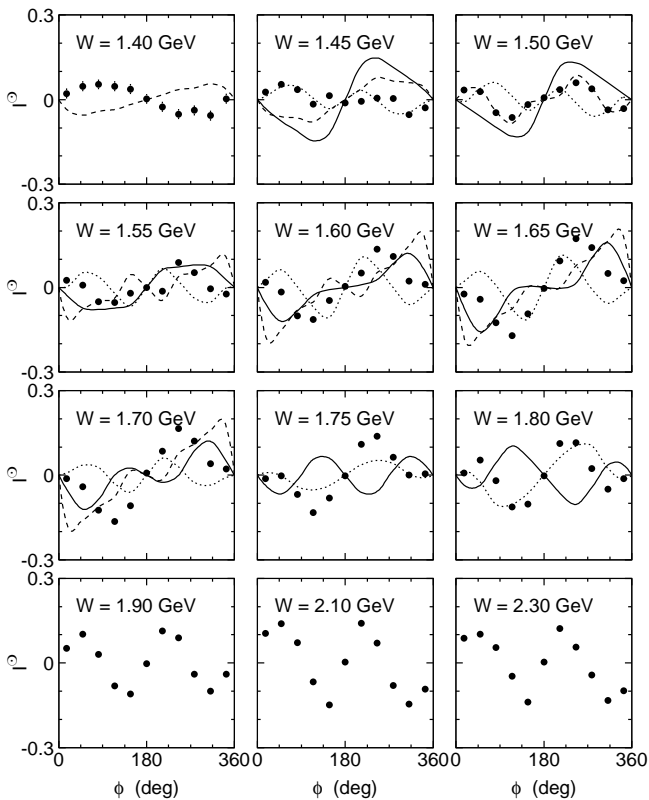}
\caption{\label{fig:all} Angular distributions for selected
center-of-mass energy bins (each with $\Delta W = 50$ MeV) of the
cross-section asymmetry for the \gp reaction. The data are integrated
over the detector acceptance. The statistical uncertainties are mostly
smaller than the symbol size. The solid and dotted curves are the
results from model calculations by Mokeev {\it et al.}
\cite{mokeev01,burkert04,mokeev04} (for $1.45\; {\rm GeV} \le W \le
1.80\; {\rm GeV}$) with relative phases of $0$ and $\pi$ between the
background- and $\pi\Delta$-subchannel amplitudes, respectively.  The
dashed curves show results of calculations by Fix and Arenh{\"o}vel
\cite{Fix04} (for $W \le 1.70$ GeV).}
\end{figure}

\begin{figure}[!ht]
\includegraphics[width=\columnwidth]{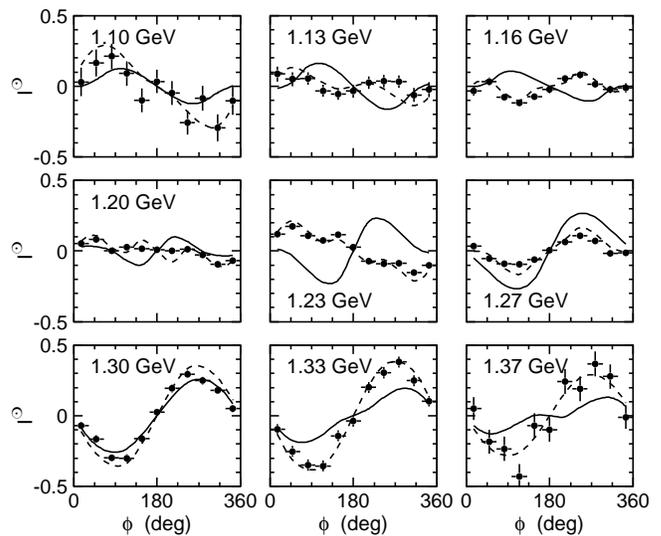}
\caption{\label{fig:d13} Helicity asymmetries at $W=1.50$~GeV for nine
bins of the invariant mass $M(p\pi^+)$, as indicated.  The solid
curves are the results of Mokeev {\it et al.}
\cite{mokeev01,burkert04,mokeev04}.  The dashed curves show
results of calculations by Fix and Arenh{\"o}vel \cite{Fix04}.}
\end{figure}

%
%
Figure \ref{fig:all} shows $\phi$ angular distributions of the
helicity asymmetry for various selected 50-MeV-wide
center-of-mass energy bins between $W = 1.40$ and $2.30$~GeV.
The data are integrated over the full CLAS acceptance. The analysis
shows large asymmetries which change markedly with $W$ up to 1.80~GeV;
thereafter they remain rather stable. The asymmetries are odd
functions of $\phi$ and vanish for coplanar kinematics ($\phi=0$ and
$180^\circ$), as expected from parity conservation \cite{Roberts04}.
The large number of observed $\vec{\gamma}p\to p\pi^+\pi^-$ events
allows for a confident analysis of the data in selected kinematic
regions, making it possible to tune the different parts of the
production amplitude independently. An example of
distributions which are more differential than those of
Fig.~\ref{fig:all} is given in Fig.~\ref{fig:d13}.  The data at
$W=1.50$~GeV are divided into nine bins in the invariant mass
$M(p\pi^+)$.

The data in Figs.~\ref{fig:all} and \ref{fig:d13} are compared with
results of available phenomenological models. In the approach by
Mokeev {\it et al.} (solid curves), double-charged-pion photo- and
electroproduction are described by a set of quasi-two-body mechanisms
with unstable particles in the intermediate states: $\pi\Delta$, $\rho
N$, $\pi N(1520)$, $\pi N(1680)$, $\pi \Delta(1600)$ and with
subsequent decays to the $\pi^+ \pi^- p$ final state
\cite{mokeev01,burkert04,mokeev04}. Residual direct $\pi^+ \pi^- p$
mechanisms are parametrized by exchange diagrams \cite{mokeev04}. The
first two quasi-two-body channels mentioned above are described by a
coherent sum of $s$-channel $N^*$ contributions and nonresonant
mechanisms \cite{mokeev01}. All well established resonances with
observed double-pion decays are included, plus $\Delta(1600)$,
$N(1700)$, $N(1710)$, and a new state, $N(1720)$ with $J^\pi={3/2}^+$,
possibly observed in CLAS double-pion data \cite{Ripani03}.  $N^*$ and
nonresonant parameters are fitted to the CLAS cross-section data for
virtual-photon double-charged-pion production \cite{Ripani03}. The
model provides a good description of all available CLAS cross-section
and world data on double-pion photo- and electroproduction at $W<1.9$
GeV and $Q^2<1.5$ GeV$^2$.

Results also have been obtained by Fix and Arenh{\"o}vel using the
model described in \cite{Fix04}.  They use an effective Lagrangian
approach with Born and resonance diagrams at the tree level. The model
includes the nucleon, the $\Delta(1232)$, $N(1440)$, $N(1520)$,
$N(1535)$, $N(1680)$, $\Delta(1620)$, $N(1675)$, and $N(1720)$
resonances, as well as the $\sigma$ and $\rho$ mesons.  The
corresponding results are shown in Figs.~\ref{fig:all} and
\ref{fig:d13} as dashed curves. For completeness, we note that the
recent work of Roca \cite{Roca04} shows our preliminary data
\cite{Strauch04} in the framework of the Valencia model for
double-pion photoproduction.

Although both models had previously provided a good description of
unpolarized cross sections, neither of the models is able to provide a
reasonable description of the beam-asymmetry data over the entire
kinematic range covered in this experiment. Even though the model
predictions agree remarkably well for certain conditions (see, {\it
e.g.}, the dashed curves in Fig.~\ref{fig:d13}), for other conditions
they are much worse and sometimes even out of phase entirely.

As is noted above, the main theoretical challenge for double-pion
photoproduction lies in the fact that several subprocesses may
contribute, even though any given individual contribution may be
small.  In this connection, the polarization measurements should be
very helpful in separating the individual terms.  The particular
sensitivity of the beam asymmetry to interference effects among
various amplitudes is illustrated in Fig.~\ref{fig:all}. The dotted
curves show results of calculations by Mokeev {\it et al.} with a
relative phase of $\pi$ between the background- and
$\pi\Delta$-subchannel amplitudes. The access to interference effects
permit a cleaner separation of background and resonances. This in term
makes it possible to make more reliable statements about the existence
and properties of nucleon resonances.

\begin{figure}[ht!]
\includegraphics[width=\columnwidth]{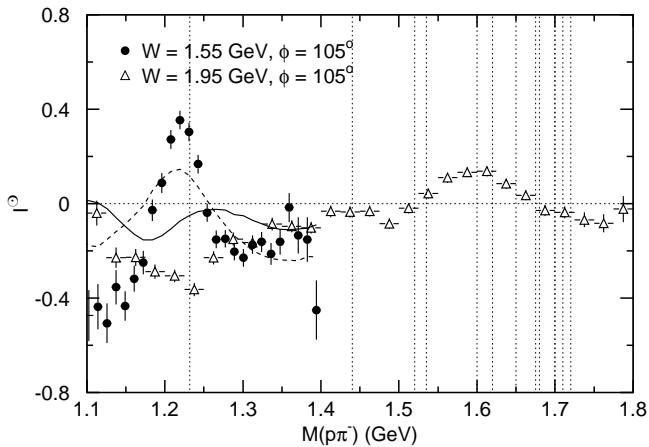}
\caption{\label{fig:phicut}Helicity asymmetry as a function of the
invariant mass $M(p\pi^-)$ for $W= 1.55$~GeV (filled circles) and
1.95~GeV (open triangles) and a $30^\circ$-wide $\phi$-angle range
centered at $\phi = 105^\circ$. The curves are the results of Mokeev
{\it et al.} \cite{mokeev01,burkert04,mokeev04} (solid) and Fix and
Arenh{\"o}vel \cite{Fix04} (dashed) for $W = 1.55$~GeV only. Note that
the result of Fix and Arenh{\"o}vel is in phase with the data (filled
circles) and that of Mokeev {\it et al.} is not. The vertical lines
indicate the masses of the known $N$ and $\Delta$ resonances.}
\end{figure}

Figure~\ref{fig:phicut} shows the helicity asymmetry as a function of
the invariant mass $M(p\pi^-)$ for two different values of $W$ and a
fixed value of $\phi$.
This is a typical case. The most interesting features of these data
are the changes that occur as $M(p\pi^-)$ traverses the $\Delta(1232)$
resonance.  At $W = 1.55$~GeV, a maximum is seen in the region of this
resonance. We see a similar trend in the region of the higher-mass
resonances around 1.60~GeV for $W=1.95$~GeV. This hints at the way in
which the helicity asymmetry (along with other polarization
observables) could be used in studies of baryon spectroscopy.  Of
particular interest is the study of sequential decays of resonances,
such as $N(1520) \to \pi \Delta \to \pi\pi N$, or $N(1700) \to \pi
N(1520) \to \pi\pi N$, which can be studied at moderate values of $W$
from 1.5 to 1.9~GeV; see \cite{Philips03}. Here, the $\rho$-production
channel is also open. This is the energy range where yet-unobserved
resonances are predicted to lie \cite{capstick94}.

In summary, we have given a brief overview of our \gp data, and we
have demonstrated, by means of phenomenological models, the
sensitivity of this helicity-asymmetry observable to the dynamics of
the reaction.  The large amount of high-quality data that we have
obtained opens the path for a series of further
investigations. Obvious next steps are (1) a better theoretical
understanding of the reaction and (2) an attempt to describe
simultaneously our polarized double-charged pion photoproduction data
and other CLAS data obtained with unpolarized real \cite{Bellis04} and
virtual \cite{Ripani03} photons.

We see, even from the small sample of data shown here, that existing
theoretical models have severe shortcomings in the description of the
beam-helicity asymmetries. In the region of overlapping nucleon
resonances (and uncontrolled backgrounds), it clearly will be a
challenge to any theoretical model to describe this new observable
that depends so sensitively on the interferences between them. Yet,
without a proper understanding of the $\pi\pi N$ channel the problem
of the ``missing'' resonances is unlikely to be resolved.

\begin{acknowledgments}
We would like to thank the staff of the Accelerator and Physics
Divisions at Jefferson Lab, as well as the Italian Istituto Nazionale
di Fisica Nucleare, the French Centre National de la Recherche
Scientifique and Commissariat \`a l'Energie Atomique, the
U.S. Department of Energy and National Science Foundation, and the
Korea Science and Engineering Foundation. Southeastern Universities
Research Association (SURA) operates the Thomas Jefferson National
Accelerator Facility under U.S. Department of Energy contract
DE-AC05-84ER40150. The GWU Experimental Nuclear Physics Group is
supported by the U.S. Department of Energy under grant
DE-FG02-95ER40901.
\end{acknowledgments}

\bibliography{g1c_helicity}

\begin{thebibliography}{30}
\bibitem[{met()}]{method}
\bibinfo{note}{The mean value $\bar I^\odot$ of model asymmetries in a
  kinematical bin is given by $\bar P_\gamma \bar I^\odot = \frac{1}{N} \sum_i
  P_{\gamma,i} I^\odot_i, $ where the sum runs over all $N$ events observed in
  that bin; $P_{\gamma,i}$ is the calculated beam polarization at the
  corresponding $W$ and $I^\odot_i$ is the model asymmetry for the kinematics
  of each of those events.}
\end{thebibliography}

\end{document}